\documentclass[aip,pop,amsmath,amssymb,reprint,superscriptaddress]{revtex4-1}

\usepackage{graphicx} 
\usepackage{dcolumn}
\usepackage{bm}
\usepackage{amssymb}
\usepackage{amsmath}
\usepackage{wasysym}
\usepackage{color}
\usepackage{soul}
\usepackage{hyperref}
\usepackage[toc,page]{appendix}
\usepackage{float}
\usepackage{flushend}
\usepackage{hyperref}
\hypersetup{
    colorlinks,%
    citecolor=blue,%
    filecolor=blue,%
    linkcolor=blue,%
    urlcolor=blue
}

\begin{document}

\preprint{AIP/123-QED}

\title{Amplification of large-scale magnetic field  in nonhelical magnetohydrodynamics}

\author{Rohit Kumar}
 \email{rohitkumar.iitk@gmail.com}
\affiliation{Institut de Recherche en Astrophysique et Plan{\'e}tologie, Universit{\'e} de Toulouse, CNRS, UPS, CNES, 31400 Toulouse, France}
 \author{Mahendra K. Verma}
\email{mkv@iitk.ac.in}
\affiliation{Department of Physics, Indian Institute of Technology, Kanpur 208016, India}


\begin{abstract} 
It is typically assumed that the kinetic and magnetic helicities play a crucial role in the growth of large-scale dynamo. In this paper  we demonstrate that helicity is not essential for the amplification of large-scale magnetic field.  For this purpose, we perform nonhelical magnetohydrodynamic (MHD) simulation, and show that the large-scale magnetic field can grow in nonhelical MHD when random external forcing is employed at scale $1/10$ the box size. The energy fluxes and shell-to-shell transfer rates computed using the numerical data show that the large-scale magnetic energy grows due to the energy transfers from the velocity field at the forcing scales.
\begin{description}
\item[PACS numbers]{47.35.Tv, 47.65.-d, 47.27.-i}
\end{description}
\end{abstract}

\maketitle

\section{introduction}
\label{sec:intro}

The generation of magnetic field in stars, planets, and galaxies is explained by dynamo effect wherein the stretching, twisting, and folding of magnetic field lines by flow generate and maintain the magnetic field.~\cite{Moffatt:book,Vainshtein:SPU1972,Childress:book} If the magnetic field is generated at the largest scales of the system, it is referred to as large-scale dynamo (LSD),~\cite{Cattaneo:JFM2002,Brandenburg:APJ2001,Brandenburg:APJ2009,Subramanian:MNRAS2014} whereas if it grows at  small scales, it is termed as small-scale dynamo (SSD).~\cite{Schekochihin:APJ2004,Schekochihin:PRL2004a,Kumar:EPL2013} Large-scale dynamos are observed in Earth, Sun, galaxies, and in experimental dynamos.~\cite{Glatzmaier:NATURE1995,Olson:JGR1999,Choudhuri:book2015,Gailitis:PRL2000,Stieglitz:PF2001,Monchaux:PRL2007}

The magnetic Prandtl number ($\mathrm{Pm}$) is an important parameter for dynamo studies. The magnetic Prandtl number for planets and stars are very small, whereas for galaxies it is very large.~\cite{Plunian:PR2013} Researchers have observed small-scale and large-scale dynamos in numerical simulations for  both small and large magnetic Prandtl numbers.~\cite{Schekochihin:APJ2004,Schekochihin:PRL2004a,Brandenburg:APJ2009,Brandenburg:APJ2014}  Note that the magnetic Prandtl number is the ratio of kinematic viscosity ($\nu$) and the magnetic diffusivity ($\eta$), and the magnetic Reynold number $\mathrm{Rm} = UL/\eta$, where $U,L$ are the large-scale velocity and length, respectively.  In a helical MHD simulation for $\mathrm{Pm} \geq 1$, Brandenburg~\cite{Brandenburg:APJ2001} observed the growth of a large-scale dynamo, which was attributed to the $\alpha$-effect in mean-field MHD. Brandenburg~\cite{Brandenburg:APJ2009} reported large-scale dynamo for $\mathrm{Pm} \leq 1$. He however argued that helicity plays a crucial role in the growth of a large-scale magnetic field. Candelaresi and Brandenburg~\cite{Candelaresi:PRE2013} also stressed  the requirement of helicity for large-scale dynamo.

Yousef {\em et al}.~\cite{Yousef:PRL2008} observed  growth of a large-scale magnetic field in a nonhelical MHD under the application of a linear shear. For fully helical flows with $\mathrm{Pm} <1$, Subramanian and Brandenburg~\cite{Subramanian:MNRAS2014} observed growth of both the small-scale and the large-scale magnetic fields. Karak {\em et al}.~\cite{Karak:APJ2015} studied large-scale dynamo transition for $\mathrm{Pm} =1$ by varying the helicity in the system and reported a critical helicity below which no large-scale field was observed. Ponty and Plunian~\cite{Ponty:PRL2011} studied the transition between large-scale and small-scale dynamos using a helical flow. For $\mathrm{Pm} <1$, as they increased $\mathrm{Rm}$, they observed the transition from large-scale dynamo to small-scale dynamo.

The growth of a large-scale dynamo is generally attributed to the $\alpha$-effect in a system.~\cite{Brandenburg:APJ2001}  Here, the growth of the magnetic field is characterized by a parameter, $\alpha$, which is proportional to the kinetic helicity.~\cite{Moffatt:book,Frisch:PD1987,Verma:PR2004,Dormy:book}  Based on the numerical simulations of rotating convective dynamo, Guervilly {\em et al}.~\cite{Guervilly:PRE2015,Guervilly:JFM2017}   proposed that large-scale vortices can generate large-scale dynamo for the magnetic Reynolds number large enough for dynamo action and small enough not to sustain small-scale magnetic field.  This is because the small-scale magnetic field suppresses the formation of large-scale vortices. Similar observations have been reported by Tobias {\em et al}.~\cite{Tobias:APJ2007} They also observed that the large-scale field is  efficiently produced for small $\mathrm{Pm}$ ($< 1$), but it is not so for $\mathrm{Pm} > 1$; for large Pm, the large-scale vortices are destroyed by the small-scale field.   

The aforementioned work appear to indicate that kinetic and magnetic helicities are important for the growth of large-scale magnetic field. Note however that R{\"a}dler and Brandenburg~\cite{Radler:PRE2008} observed large-scale magnetic field generated through  $\alpha$-effect in mean-field dynamo theory, where the kinetic helicity was zero.  In this paper we  show that helicity is not essential for the above process.  We perform direct numerical simulation of  nonhelical MHD with forcing in the intermediate regime ($k=[10, 12]$, between small-$k$ modes corresponding to the box size and the dissipative large-$k$ modes), and demonstrate that the large-scale dynamo occurs due to the energy transfers from the velocity field at forcing-scale to the large-scale magnetic field. The energy transfers are quantified using the method proposed by Dar {\em et al.}~\cite{Dar:PD2001} and Verma.~\cite{Verma:PR2004} We however remark that an injection of appropriate kinetic and magnetic helicities may increase the growth rate of large-scale magnetic field.~\cite{Verma:PR2004,Pouquet:JFM1976,Verma:Pramana2003b}

Kumar {\em et al.}~\cite{Kumar:EPL2013} and Kumar {\em et al.}~\cite{Kumar:JT2015}   computed the energy transfers in small-scale dynamo and large-scale dynamo using similar scheme, but the forcing in their simulations was at large scales. In the large-scale dynamo of Kumar {\em et al},~\cite{Kumar:JT2015} the dominant energy transfer to the magnetic field was at the forcing length scale of the velocity field, which was the largest length scale of the system. Debliquy {\em et al}.~\cite{Debliquy:PP2005} studied energy transfers in decaying MHD for the unit $\mathrm{Pm}$. They used logarithmically binned wavenumber shells to compute various energy fluxes and shell-to-shell energy transfers, similar to that of Dar {\em et al}.~\cite{Dar:PD2001} On the other hand, Alexakis {\em et al}.~\cite{Alexakis:PRE2005} used linearly binned shells to quantify the energy transfers in MHD and reported local transfers between the same fields, but velocity to magnetic transfers were predominantly nonlocal. Moll {\em et al.}~\cite{Moll:APJ2011} employed Alexakis {\em et al.}'s method~\cite{Alexakis:PRE2005} and quantified the shell-to-shell energy transfers in a small-scale dynamo. They observed that during the dynamo growth, energy transfers take place from  large-scale velocity field to  small-scale magnetic field. Note that Alexakis {\em et al}.~\cite{Alexakis:PRE2005}, Moll {\em et al.}~\cite{Moll:APJ2011}, Kumar {\em et al.}~\cite{Kumar:EPL2013}, and Kumar {\em et al.}~\cite{Kumar:JT2015} forced the large-scale velocity field, whereas in this paper, the forcing is in the intermediate-scale. We show that this kind of forcing allows the magnetic field to grow at large scales.

The paper is organized as follows: In Sec.~\ref{sec:theory}, we present the governing MHD equations and the formalism for the calculation of energy fluxes and shell-to-shell energy transfer rates. Details of numerical simulation are presented in Sec.~\ref{sec:simulation}. In Sec.~\ref{sec:results}, we present results of the forced MHD simulation for $\mathrm{Pm} =1$. Finally, in Sec.~\ref{sec:conclude} we summarize our simulation results.     
     
\section{Formulation of the problem} 
\label{sec:theory}
The governing equations for the dynamo process are~\cite{Verma:PR2004}
\begin{eqnarray}
\partial_{t}\mathbf{u}+ (\mathbf{u} \cdot \nabla) \mathbf{u} & = & -\nabla \left(\frac{p}{\rho}\right)+ \frac{\mathbf{J} \times \mathbf{b}}{\rho}
+ \nu \nabla^{2}\mathbf{u} + \mathbf{F}, \label{eq:MHD_vel}\\
\partial_{t}\mathbf{b}+ (\mathbf{u} \cdot \nabla) \mathbf{b} & = & (\mathbf{b} \cdot \nabla) \mathbf{u}+ \eta \nabla^{2}\mathbf{b}, \label{eq:MHD_mag}\\
\nabla \cdot \mathbf{u} & = & 0, \\ \label{eq:div_u_0}
\nabla \cdot \mathbf{b} & = & 0,  \label{eq:div_b_0}
\end{eqnarray}
where $\mathbf{u}$ is the velocity field, $\mathbf{b}$ is the magnetic field, $\mathbf{J} = \nabla \times \mathbf b$ is the current density, $p$ is the thermal pressure, $\rho$ is the  fluid density, and $\mathbf{F}$ is the external force field.   We assume the flow to be incompressible, and choose the density to unity. In this paper we solve the above equations using computer simulation, and show that the large-scale magnetic field grows due to energy transfers to it from the velocity field at the forcing scales.

In order to quantify the energy transfers between velocity and magnetic fields, we compute energy fluxes and shell-to-shell energy transfer rates using the formulation of Dar {\em et al}.\cite{Dar:PD2001}, Verma,~\cite{Verma:PR2004} and Debliquy {\em et al}.~\cite{Debliquy:PP2005} The energy flux from the region $X$ (of wavenumber space) of $\mu$ field  to the region  $Y$ of $\beta$ field is defined as  
\begin{eqnarray}
\Pi^{\mu,X}_{\beta,Y} = \displaystyle\sum_{\mathbf{k} \in Y} \displaystyle\sum_{\mathbf{p} \in X} S^{\beta \mu} (\mathbf{k} | \mathbf{p} |\mathbf{q}), \label{eq:flux}
\end{eqnarray}
where $S^{\beta \mu} (\mathbf{k}| \mathbf{p}| \mathbf{q})$ is the energy transfer rate from mode $\mathbf{p}$ of $\mu$ field to mode $\mathbf{k}$ of $\beta$ field, where the mode $\mathbf{q}$ acts as a mediator.  The triadic modes ($ {\mathbf {k,p,q}}$) satisfy a condition  $\mathbf{k} + \mathbf{p} + \mathbf{q} = 0$. As an example, the energy transfer rate from $\mathbf{u} (\mathbf{p})$ to $\mathbf{b} (\mathbf{k})$ is~\cite{Dar:PD2001} 
\begin{eqnarray}
S^{bu} (\mathbf{k} | \mathbf{p} | \mathbf{q}) = \Im ([\mathbf{k} \cdot \mathbf{b} (\mathbf{q})][\mathbf{b} (\mathbf{k}) \cdot \mathbf{u} (\mathbf{p})]),\label{eq:mode-to-mode}
\end{eqnarray}
where $\Im$ denotes the imaginary part of the argument, and $\mathbf{b} (\mathbf{q})$ acts as a mediator.

\begin{figure}[htbp]
\centering
\includegraphics[width=3.50cm,angle=0]{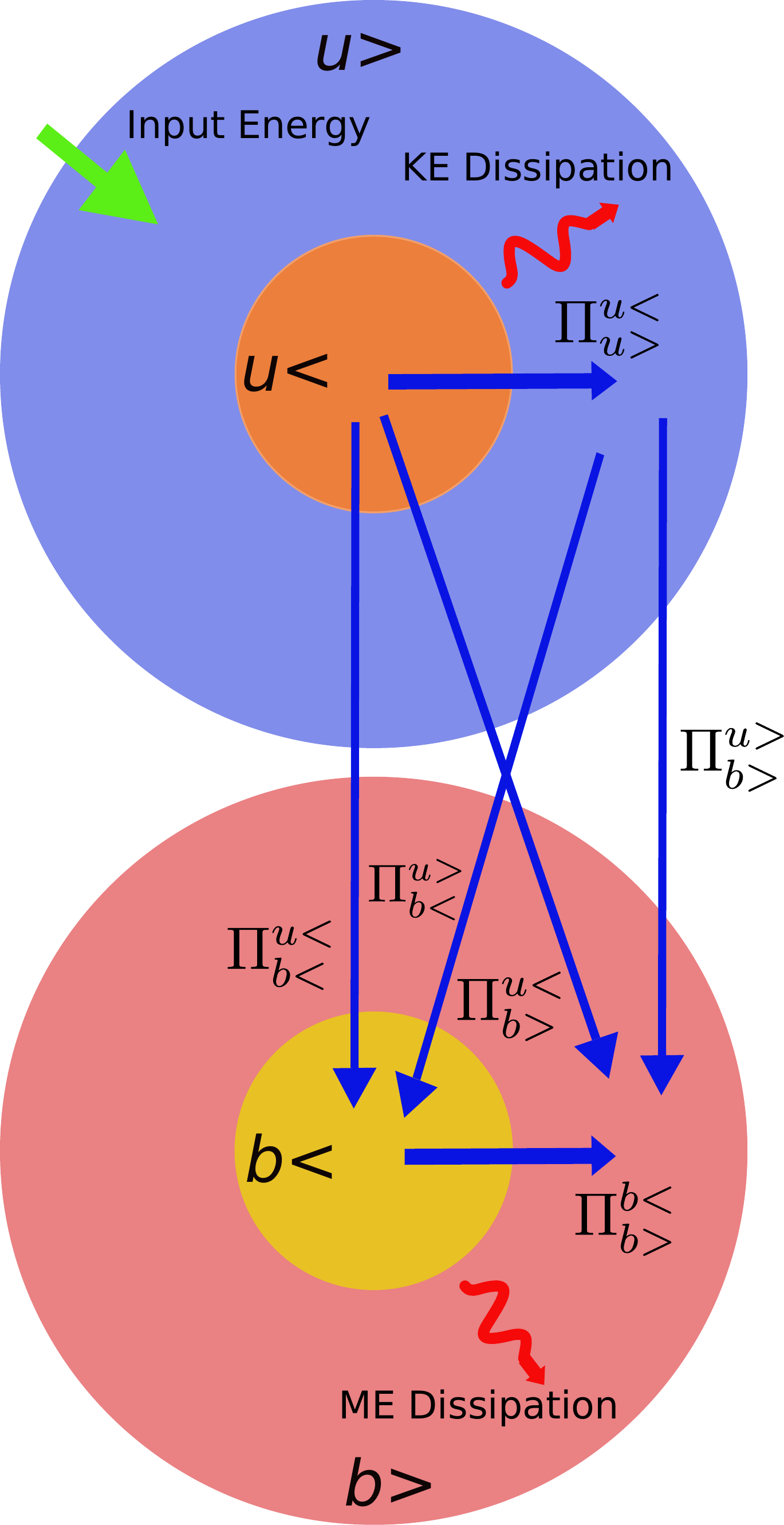} 
\caption{A schematic diagram depicting the energy fluxes, and the kinetic and the magnetic energy dissipation rates in MHD turbulence.}
\label{fig:flux_draw}
\end{figure}

 The energy fluxes in MHD turbulence are: $\Pi^{u<}_{u>}(k_0)$, $\Pi^{u>}_{b<}(k_0)$, $\Pi^{b<}_{b>}(k_0)$, $\Pi^{u<}_{b>}(k_0)$, $\Pi^{u<}_{b<}(k_0)$, and $\Pi^{u>}_{b>}(k_0)$. Here $<$ and $>$ represent the modes residing inside and outside the sphere of radius $k_0$, respectively.   A schematic representation of the above energy fluxes, and the kinetic and the magnetic energy dissipation rates are illustrated in Fig.~\ref{fig:flux_draw}.   In particular, the energy flux from outside of the $u$-sphere of radius $k_0$ to inside of the $b$-sphere of the same radius is defined as  
\begin{eqnarray}
\Pi^{u>}_{b<}(k_0) = \displaystyle\sum_{|\mathbf{k}| < k_0} \displaystyle\sum_{|\mathbf{p}| > k_0} S^{b u} (\mathbf{k} | \mathbf{p} | \mathbf{q}). \label{eq:flux_ub}
\end{eqnarray}

For MHD turbulence, Dar {\em et al}.\cite{Dar:PD2001} also formulated the shell-to-shell energy transfer rates, which facilitate us with a refined picture of the energy transfers in wavenumber space. In MHD, there are three kinds of shell-to-shell energy transfer rates\cite{Dar:PD2001,Verma:PR2004}:  from velocity to velocity field ($U2U$), from magnetic to magnetic ($B2B$), and from velocity to magnetic ($U2B$). The shell-to-shell energy transfer from the $m$-th shell of $\mu$ field to the $n$-th shell of $\beta$ field is defined as\cite{Debliquy:PP2005, Dar:PD2001, Verma:PR2004} 
\begin{eqnarray}
T^{\beta,\mu}_{n,m} = \displaystyle\sum_{\mathbf{k} \in n} \displaystyle\sum_{\mathbf{p} \in m} S^{\beta \mu} (\mathbf{k} | \mathbf{p} | \mathbf{q}). \label{eq:shell-to-shell}
\end{eqnarray}
For example, the shell-to-shell energy transfer rate from the $m$-th shell of $u$ field to the $n$-th shell of $b$ field is  
\begin{eqnarray}
T^{b,u}_{n,m} = \displaystyle\sum_{\mathbf{k} \in n} \displaystyle\sum_{\mathbf{p} \in m} S^{b u} (\mathbf{k} | \mathbf{p} | \mathbf{q}). \label{eq:shell-to-shell_ub}
\end{eqnarray}

Kumar {\em et al}~\cite{Kumar:EPL2013} used the above shell-to-shell transfer computation scheme to study the energy transfers in small-scale dynamo with $\mathrm{Pm} =20$. Kumar {\em et al}~\cite{Kumar:JT2015} carried out similar studies in a dynamo with $\mathrm{Pm} =0.2$. In both these cases the velocity field was forced at large scales, and the growth of magnetic field was observed at small and intermediate scales. In this paper, however, we force the velocity field at intermediate scales so that the growth of a large-scale magnetic field could be observed. 

In the next section, we will discuss the numerical framework for the dynamo simulation.

\section{Details of numerical simulation} 
\label{sec:simulation}

Using a pseudo-spectral code {\em Tarang},\cite{Verma:Pramana2013} we solve Eqs.~(\ref{eq:MHD_vel}-\ref{eq:div_b_0}) in a three-dimensional box of size $(2\pi)^3$ with periodic boundary conditions in all the three directions. The grid size for our simulation is $N^3$ with $N=512$. We employ Runge-Kutta fourth order (RK4) scheme for time stepping, $2/3$ rule for dealiasing, and CFL criterion for choosing $\Delta t$.  We nondimensionalize the velocity field and the magnetic field (in Alfv\'{e}nic units) using the velocity scale $U$, and position vector using the length scale $L$, and time using $L/U$. Hence, the time unit in our simulation is the eddy turnover time $L/U$. In our simulation, we choose $\nu=\eta = 5\times 10^{-3}$, thus $\mathrm{Pm}=1$.    

We employ a nonhelical random forcing to the velocity field in a wavenumber band $k =[10, 12]$ such that the kinetic energy supply rate $\epsilon$ is a constant, and is equal to unity. The force field is defined as~\cite{Carati:PF1995}  
\begin{equation}
\mathbf{F}(\mathbf{k}) = A(\mathbf{k}) \mathbf{u}(\mathbf{k}) + B(\mathbf{k}) \boldsymbol{\omega}(\mathbf{k}), \label{eq:forcing}
\end{equation}
where $\boldsymbol{\omega}(\mathbf{k}) = \nabla \times \mathbf{u}(\mathbf{k})$, and $A(\mathbf{k})$ and $B(\mathbf{k})$ are the real coefficients. Note that the forcing is employed at an intermediate scale, which is between the large scales ($k \approx 1$) and the dissipative scale ($k \approx N/2$).  The energy feed to the system due the above forcing is 
\begin{eqnarray}
\mathcal{F}(\mathbf{k}) & = & \Re[\mathbf{u}(\mathbf{k}) \cdot \mathbf{F}^*(\mathbf{k}))] \nonumber \\ 
& = &  2 A(\mathbf{k}) E_u(\mathbf{k}) + 2 B(\mathbf{k}) H(\mathbf{k}), \label{eq:en_supply}
\end{eqnarray}
where $E_u(\mathbf{k}) = \dfrac{1}{2} [\mathbf{u}(\mathbf{k}) \cdot \mathbf{u}^*(\mathbf{k})]$ is the kinetic energy, $H(\mathbf{k}) = \dfrac{1}{2} \Re[\mathbf{u}^*(\mathbf{k}) \cdot \boldsymbol{\omega}(\mathbf{k})]$ is the kinetic helicity,  $\Re$ is real part of the argument, and $\sum_{\bf k} \mathcal{F}(\mathbf{k}) = \epsilon$, the total kinetic-energy supply rate.  We take $B(\mathbf{k}) =0$  ensuring that the supply of kinetic helicity vanishes.  We only supply constant kinetic energy.  As a result,  in our simulation, the total kinetic helicity, $\int ({\bf u} \cdot \boldsymbol{\omega})/2 d{\bf r}$ ($d{\bf r}$ is the volume element), and magnetic helicity, $\int ({\bf a \cdot b})/2 d{\bf r}$ (${\bf a}$ is the vector potential), are of the order of $10^{-2}$, which are much  smaller than the total kinetic energy, $\int (u^2/2) d{\bf r}$, and the total magnetic energy, $\int (b^2/2) d{\bf r}$, which are of the order of unity in the steady state.

For the initial conditions, we employ a random velocity field at all the wavenumbers, and the seed magnetic field  at large wavenumbers ($k \ge 10$). The initial magnetic field is applied only at large wavenumbers in order to test the growth of magnetic field at small wavenumber ($k < 10$).  The simulation is carried out for till $t=t_\mathrm{final}=127$ time units (eddy turnover times).   During the final stages, the Reynolds number $\mathrm{Re} = UL/\nu \approx100$.  Note that   $\mathrm{Rm} =\mathrm{Re}$.

In the present paper, our main focus is to compute the energy fluxes and shell-to-shell energy transfers during the magnetic energy growth.  For the flux computations, we construct wavenumber spheres with their centre at the origin and their radii as $2.0$, $4.0$, $8.0$, $9.6$, $11.6$, $13.9$, $16.8$, $20.2$, $24.3$, $29.2$, $35.1$, $42.2$, $50.8$, $61.1$, $73.5$, $88.4$, $106.4$, $128.0$, and $256.0$.  For the computation of the shell-to-shell energy transfers, we divide the Fourier space   into $19$  shells with the shell centre at the origin.  The inner and outer radii of the $n$th shell are $k_{n-1}$ and $k_n$, respectively. The aforementioned radii are used for the construction of wavenumber shells. The shells in the inertial range are logarithmically binned keeping in mind the power law physics here.

In the next section, we discuss our numerical results.

\section{Results}
\label{sec:results}

As described in the previous section, we force the velocity field randomly in the wavenumber band $k=[10,12]$ and show that the magnetic energy at large-scales ($k\approx 1$) grows due to a complex energy transfer.   We carry out our simulation  till $t=t_\mathrm{final}=127$ eddy turnover times at which time $\mathrm{Re} = \mathrm{Rm} \approx 100$.  

\begin{figure}[htbp]
\centering
\includegraphics[width=8.5cm,angle=0]{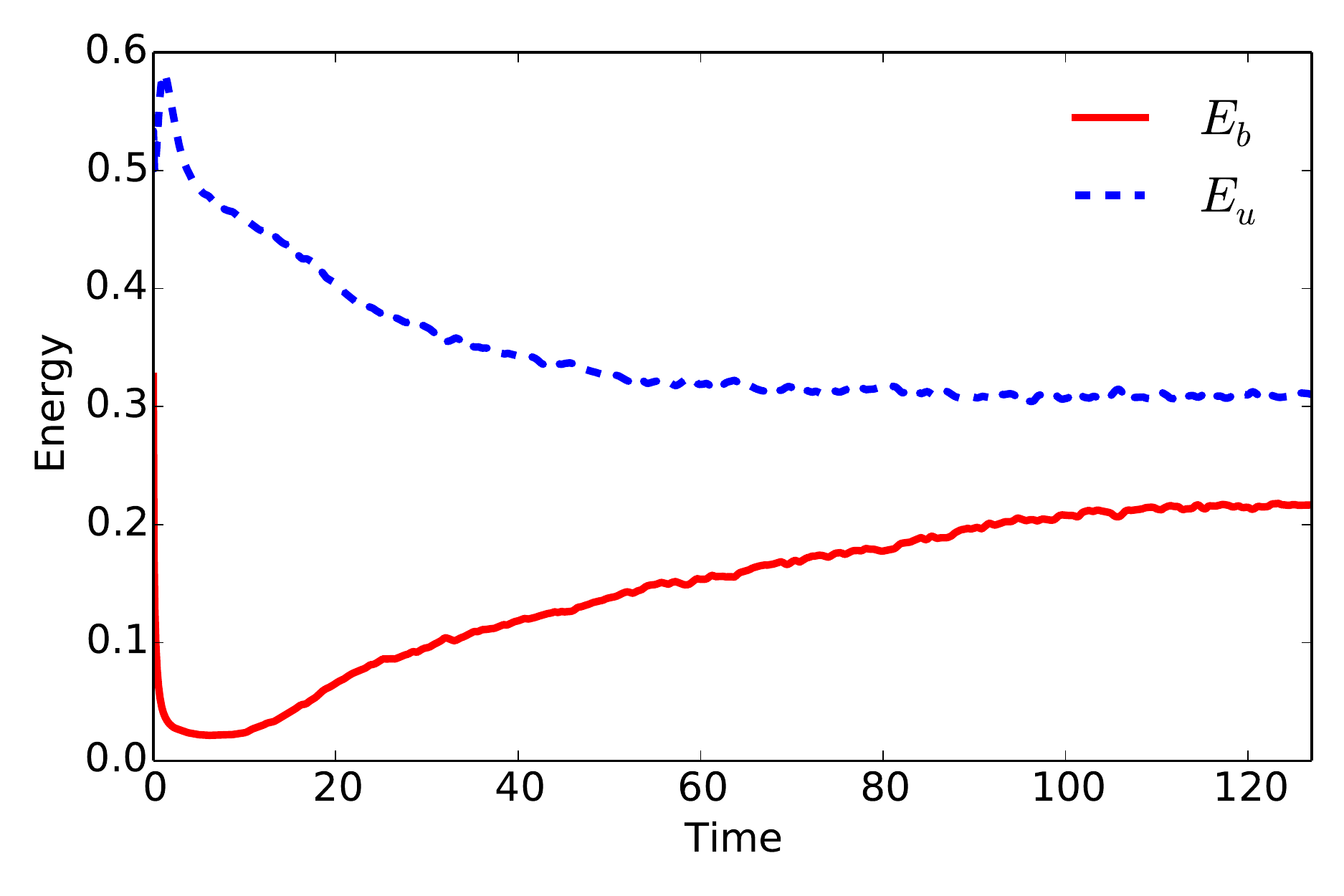} 
\caption{Evolution of kinetic energy ($E_u$) and magnetic energy ($E_b$) with time. In the saturated state $E_u/E_b \approx 1.4$.}
\label{fig:lsd_kin_mag}
\end{figure}

We present the time-evolution of the total kinetic and the total magnetic energies in Fig.~\ref{fig:lsd_kin_mag}.    In the early stages of simulation, the magnetic energy decreases quickly from its initial value; this is a transient phenomena.  The mechanical energy cascades to small scales where it is dissipated by Joule heating. After $t \approx 5$, the magnetic energy grows with time, as observed in the dynamo simulations.  Finally the kinetic and magnetic energies attain saturation near $t\approx 100$. During the saturation, the ratio   $E_u/E_b \approx 1.4$, which is close to an equipartition, as reported in many numerical simulations.~\cite{Chou:APJ2001b,Maron:APJ2004,Ponty:PRL2005,Schekochihin:APJ2004,Kumar:EPL2013} It has been observed in many numerical simulations that the magnetic energy tends to saturate in $30 - 50$ time units.~\cite{Schekochihin:APJ2004,Cattaneo:JFM2002,Chou:APJ2001b}  However, saturation in our simulation occurs around $100$ time units. We did not continue the simulation further, because our study is mainly focused on energy transfers during the large-scale magnetic field growth phase.

\begin{figure}[htbp]
\centering
\includegraphics[width=8.0cm,angle=0]{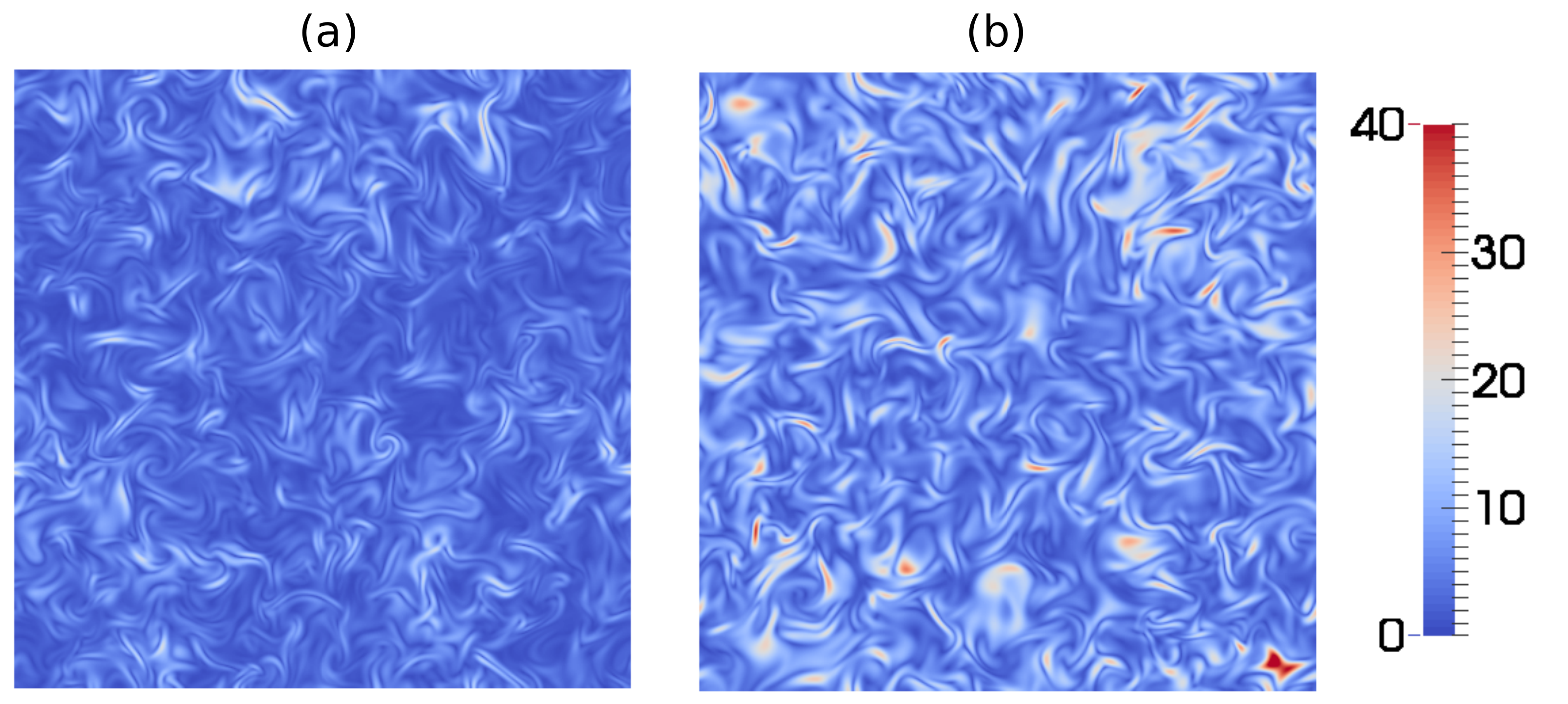} 
\caption{Density plots of current density $|\nabla \times {\mathbf b}|$ for a cross-section of the cube (a) at $t=5$ and (b) at $t=127$. In the early stages of dynamo, strong currents are present at small scales, while relatively large-scale currents are observed in the final stage.}
\label{fig:curr_den}
\end{figure}

\begin{figure}[H]
\centering
\includegraphics[width=8.0cm,angle=0]{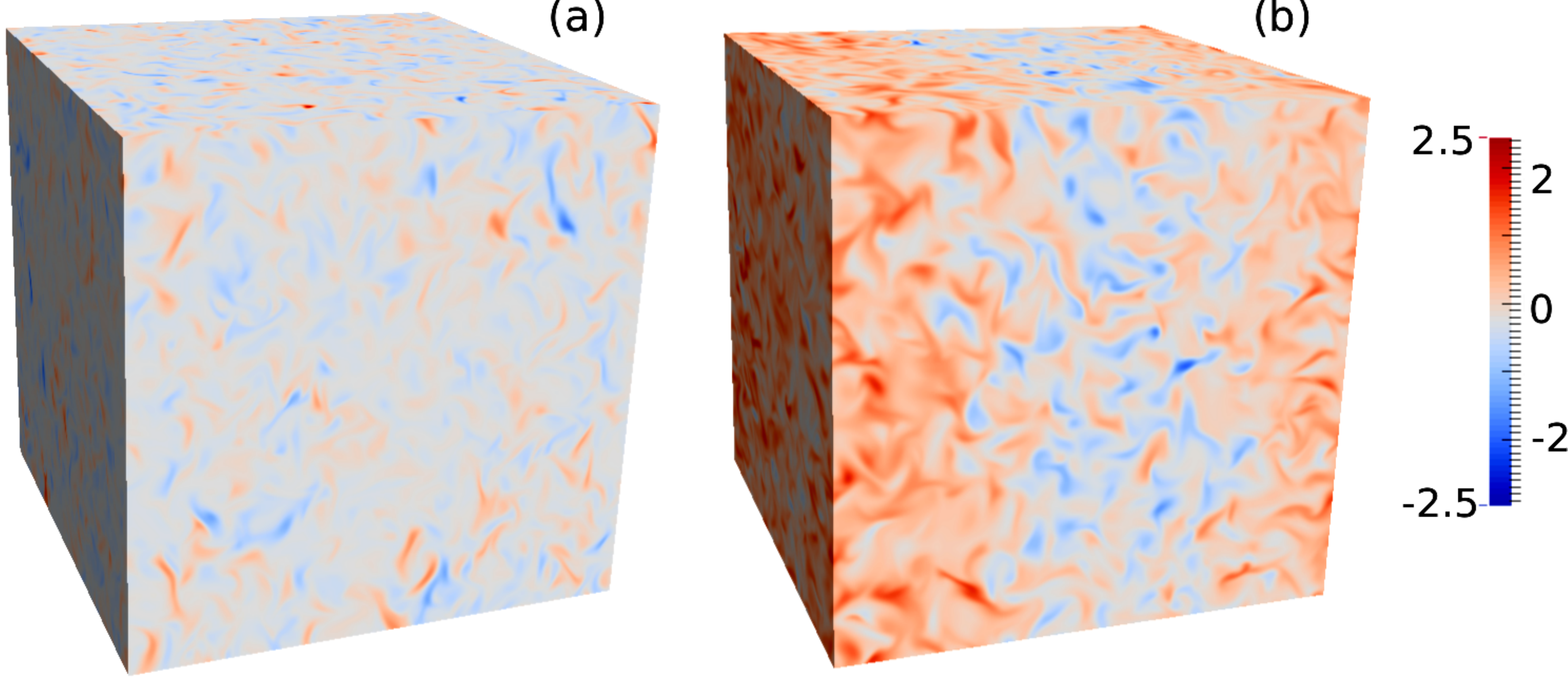} 
\caption{The y-component of the magnetic field ($B_y$) in physical space at (a) $t=5$ and (b) $t=127$. The magnetic field has large-scale structures at $t=127$.}
\label{fig:By_real}.
\end{figure}

In Fig.~\ref{fig:curr_den}, we exhibit the density plot of the current density $| \nabla \times {\mathbf b}|$ at $t=5$ and at $t=127$ (the final stage of the simulation). The initial current density at $t=5$ shows small islands of intense  $|{\mathbf J}|$ [Fig.~\ref{fig:curr_den}(a)], while the plot for final stage dynamo shows large islands of intense $|{\mathbf J}|$ [Fig.~\ref{fig:curr_den}(b)]. We also plot the y-component of the magnetic field, $B_y$, at initial and final stages of the simulation (shown in Fig.~\ref{fig:By_real}). Initially the magnetic field has small-scale structures, but in the later stages magnetic field appears with large-scale structures. These figures demonstrate growth of large-scale magnetic field. 

\begin{figure}[htbp]
\centering
\includegraphics[width=8.5cm,angle=0]{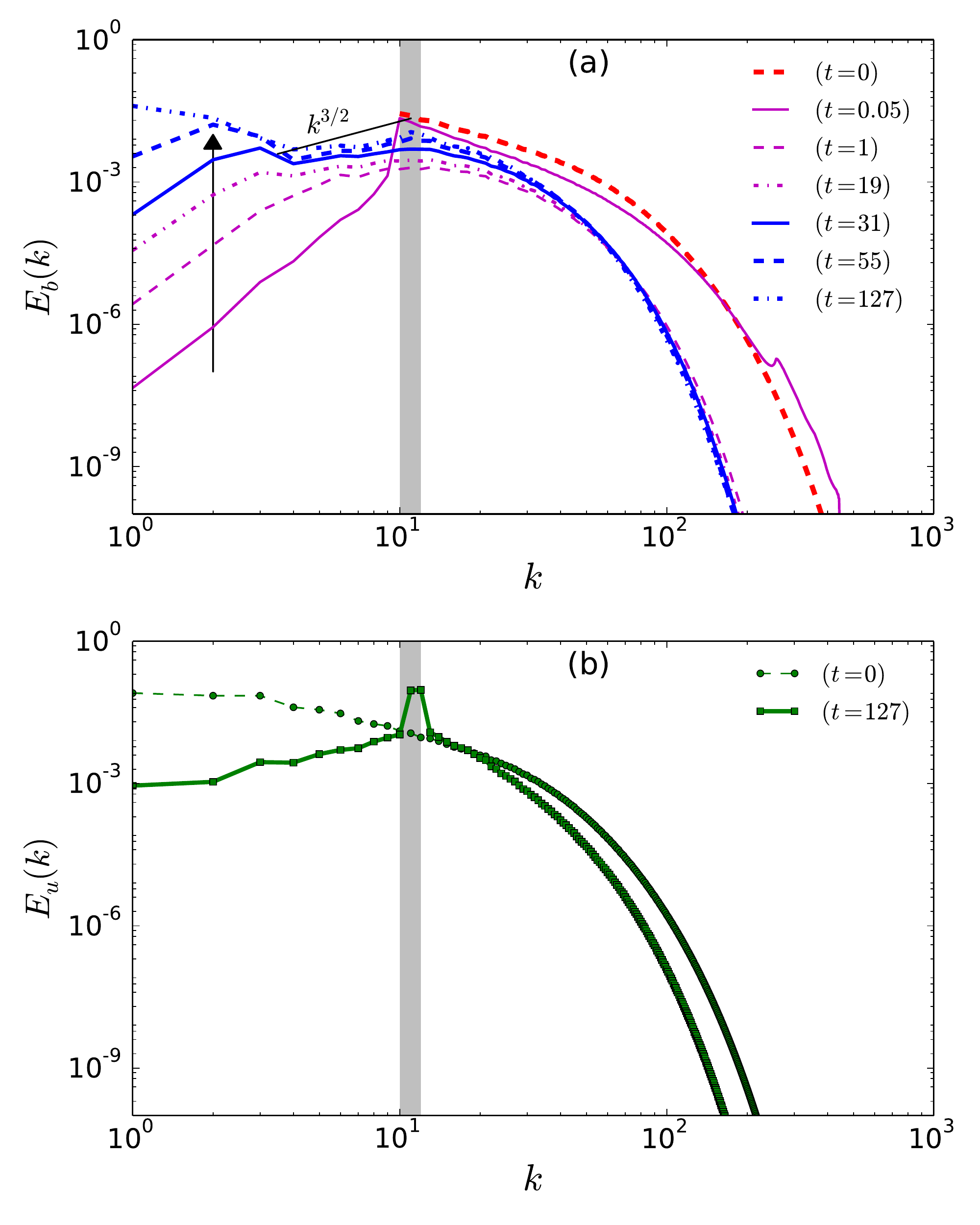} 
\caption{(a) The magnetic energy spectrum ($E_b(k)$) and (b) kinetic energy spectrum ($E_u(k)$) at various times. The growth of magnetic energy takes place at small wavenumbers or large scales.  The velocity field is forced at wavenumber band $k=[10,12]$, shown as the grey band in the figures. }
\label{fig:Ebk_LS}
\end{figure}

To quantify the evolution of magnetic and kinetic energies at different scales, we plot time-evolution of magnetic ($E_b(k)$) and kinetic ($E_u(k)$) energy spectra, which are shown in Fig.~\ref{fig:Ebk_LS}(a,b), respectively. As we have discussed in the previous section, the initial stage magnetic energy at $t=0$ is present only at the wavenumbers $k \geq 10$, i.e., at small and intermediate scales of the system. In the figure, the forcing wavenumber band $k=[10, 12]$ is shown by a shaded region. As the simulation progresses, the magnetic energy  grows at smaller wavenumbers ($k < 10$) or large length scales, as shown in Fig.~\ref{fig:Ebk_LS}(a). The growth of magnetic energy at $k=1, 2$ (large scales) demonstrate large-scale dynamo. 

\begin{figure}[htbp]
\centering
\includegraphics[width=8.2cm,angle=0]{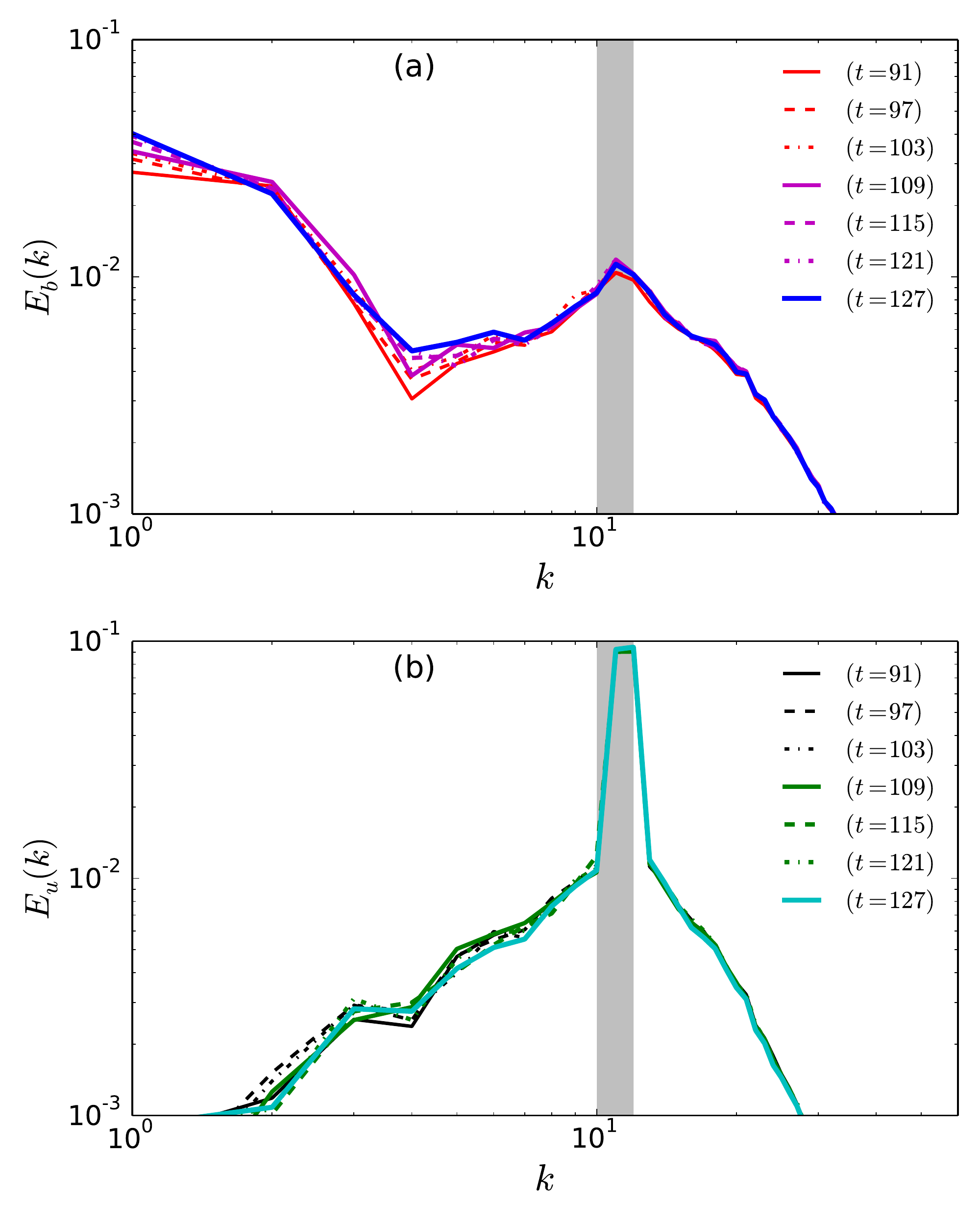} 
\caption{(a) The magnetic energy spectrum ($E_b(k)$) and (b) kinetic energy spectrum ($E_u(k)$) during the final stage when saturation has taken place. The grey strip indicates the forcing wavenumber band $k=[10,12]$, as illustrated in Fig.~\ref{fig:Ebk_LS}.}
\label{fig:Ek_final}
\end{figure}

\begin{figure}[htbp]
\centering
\includegraphics[width=8.5cm,angle=0]{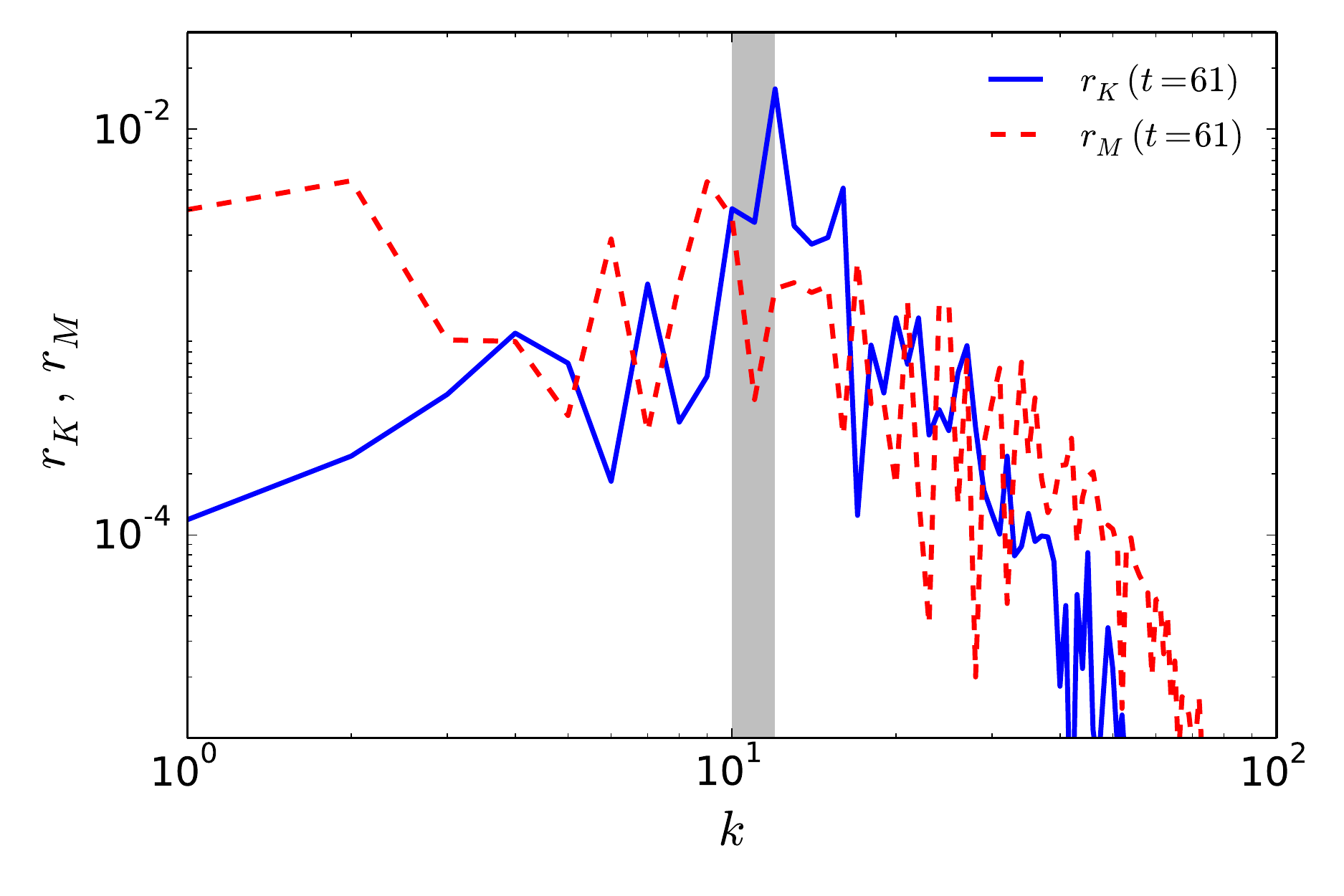} 
\caption{At $t=61$, the plots of $r_K = |H_K(k)|/(k E_u(k))$ and $r_M = k |H_M(k)|/E_b(k)$ with $H_K(k), H_M(k)$ as the kinetic helicity and the magnetic helicity spectra, respectively. The grey strip indicates the forcing wavenumber band $k=[10,12]$. $r_K, r_M \ll 1$ indicate that helicities are small in our simulations.}
\label{fig:rK_rM}
\end{figure}

In Fig.~\ref{fig:Ek_final}, we exhibit the magnetic and kinetic energy spectra for $t=91$ to 127 time units. The figure clearly demonstrates that the energy spectra have reached saturation. In Fig.~\ref{fig:rK_rM} we plot $r_K = |H_K(k)|/(k E_u(k))$ and $r_M = k |H_M(k)|/E_b(k)$ for $t=61$, which is intermediate time before saturation.  Here $H_K(k), H_M(k)$ are the kinetic helicity and the magnetic helicity spectra, respectively.  The smallness of $r_K$ and $r_M$ indicates that the helicities are negligible in our simulation.

\begin{figure}[htbp]
\centering
\includegraphics[width=8.0cm,angle=0]{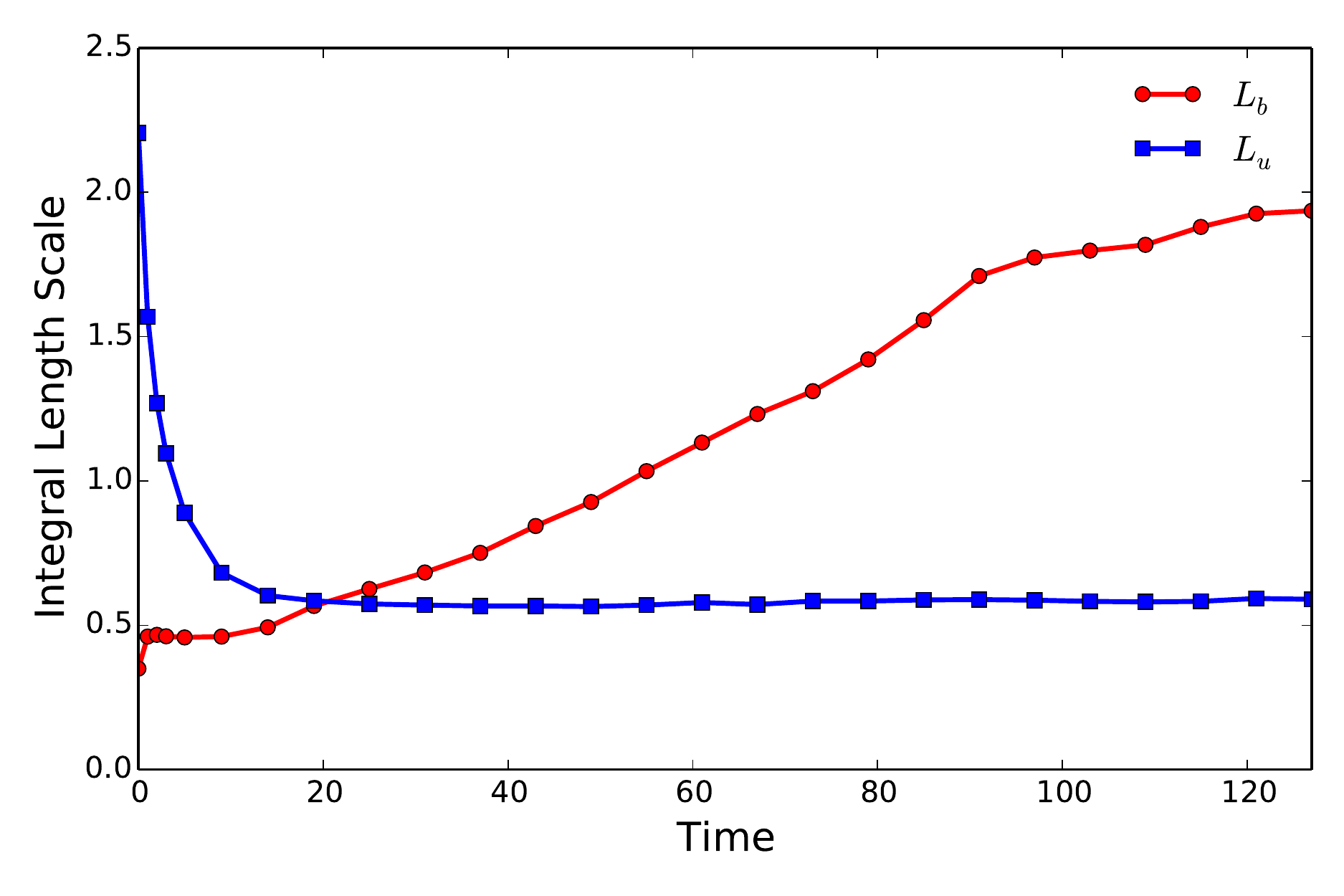} 
\caption{Time-evolution of velocity and magnetic integral length scales ($L_u, L_b$). In the final stages $L_b/L_u \approx 3$.}
\label{fig:lsd_integ_len_time}
\end{figure}

We compute velocity and magnetic integral length scales, which are defined as   $L_u =2\pi \int k^{-1} E_u(k) dk / \int E_u(k) dk$ and $L_b =2\pi \int k^{-1} E_b(k) dk / \int E_b(k) dk$, respectively. We plot $L_u$ and $L_b$ in Fig.~\ref{fig:lsd_integ_len_time}. At $t=0$ the magnetic energy is concentrated only at small and intermediate scales, hence $L_b$ is smaller than $L_u$. But at later times during the magnetic energy growth, $L_b$ quickly grows larger than $L_u$. In the final stages $L_b/L_u \approx 3$, which again corroborates the growth of a large-scale magnetic field in the system.      

In the next subsections we will focus on energy transfers between velocity and magnetic fields during the magnetic energy growth.

\subsection{Energy Fluxes}

\begin{figure}[htbp]
\centering
\includegraphics[width=7.5cm,angle=0]{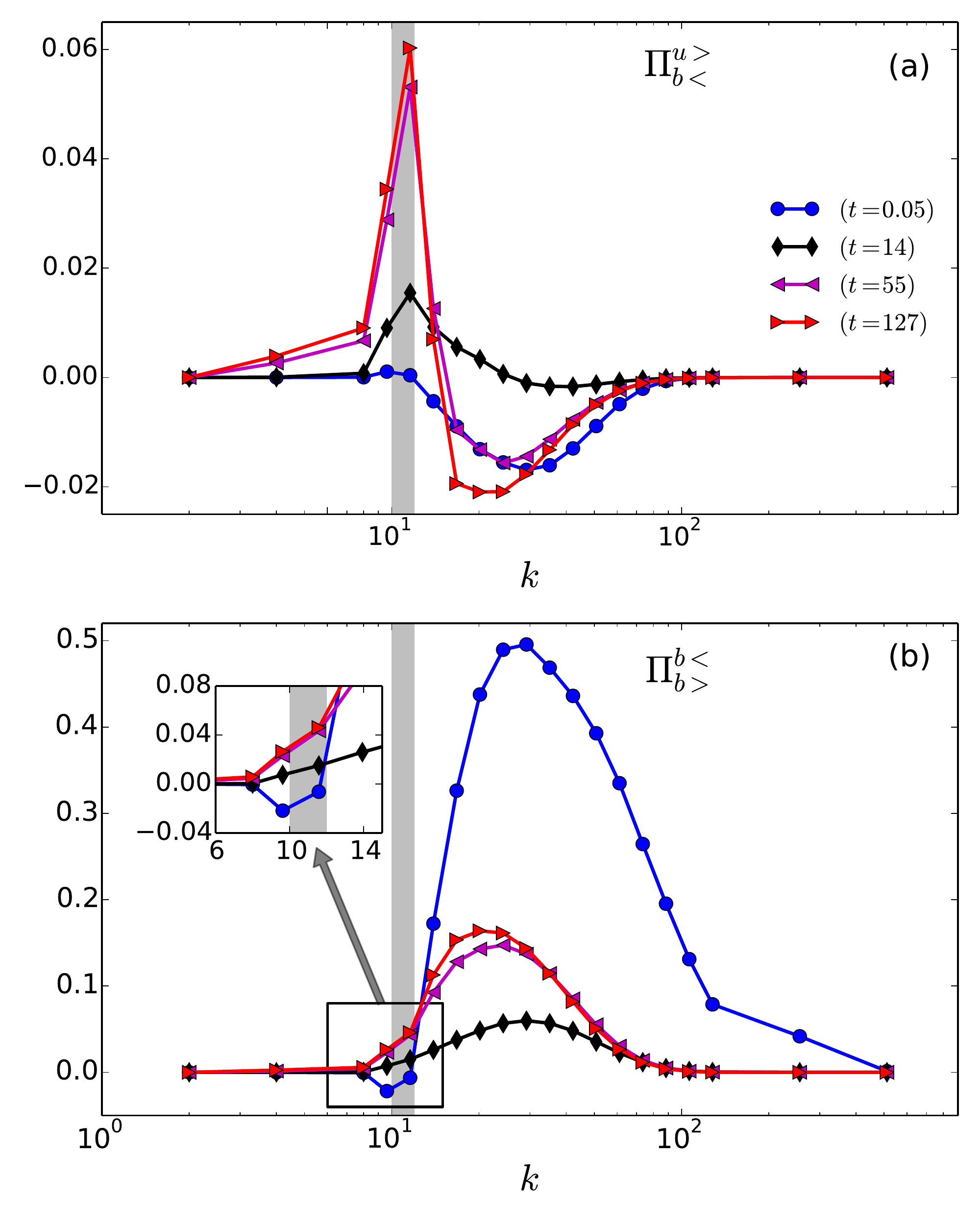} 
\caption{Plots of energy fluxes $\Pi^{u>}_{b<}(k)$ and $\Pi^{b<}_{b>}(k)$ vs.~$k$. The inset exhibits a zoomed view of the energy flux near the forcing wavenumber range. The shaded regions represent the forcing wavenumber band.}
\label{fig:lsd_fluxes}
\end{figure}

In this section, we compute energy fluxes of MHD turbulence for  wavenumber spheres with radii mentioned in Sec.~\ref{sec:simulation}. We perform these computations at different times.  In Fig.~\ref{fig:lsd_fluxes} we illustrate $\Pi^{u>}_{b<}$ ($u>$ to $b<$) and $\Pi^{b<}_{b>}$ ($b<$ to $b>$) for various spheres.  For the following discussion, it is important to keep in mind that the forcing wavenumber band $k=$(10--12), shown as the grey strip in the figure, lies beyond the fourth sphere, but inside the fifth sphere.

Figure~\ref{fig:lsd_fluxes}(a) indicates that  $\Pi^{u>}_{b<} (k\le 9) >0$ throughout the simulation.  Also, $\Pi^{u>}_{b<}$ is most dominant for $k \approx 10$ indicating a local energy transfer.  This feature indicates that the large-scale magnetic field receives energy from the velocity field $u>$, which is one of the prime sources for the large-scale dynamo.  As shown in Fig.~\ref{fig:lsd_fluxes}(b), at $t=0.05$,  $\Pi^{b<}_{b>}(k=9) < 0$ indicating an inverse cascade of magnetic energy at early times.  These energy transfers strengthen the large-scale magnetic field.  This energy transfer is reminiscent of quick spread of energy to larger wavenumbers in hydrodynamic turbulence.  Here in MHD turbulence we observe that the small wavenumber magnetic modes receive energy from the velocity and magnetic modes ($u>, b>$) due to nonlinear interactions.  

As the magnetic energy at $k < 10$ grows after initial transients, the energy flux $\Pi^{b<}_{b>}$ becomes positive, indicating a forward cascade of magnetic energy.  This result is consistent with $\Pi^{b<}_{b>}$ computations under steady state.~\cite{Debliquy:PP2005,Kumar:EPL2013} The energy flux $\Pi^{u>}_{b<}$ is positive for $k< 10$ that yields steady growth of large-scale magnetic field till saturation.

\begin{figure}[htbp]
\centering
\includegraphics[width=6.0cm,angle=0]{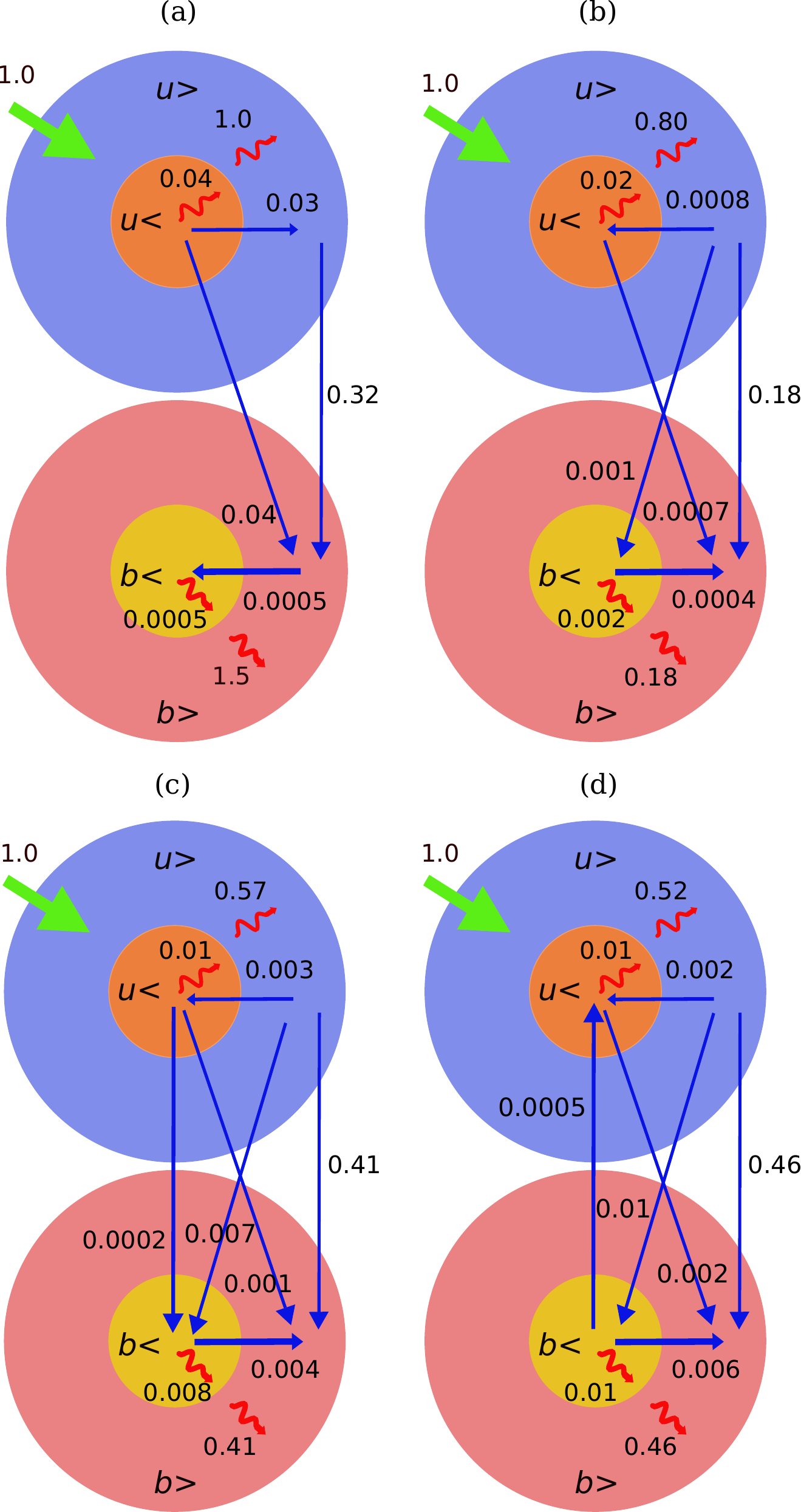} 
\caption{Schematic diagrams representing the energy fluxes and dissipation rates for wavenumber sphere $k=8$ at different times: (a) $t=0.05$, (b) $t=14$, (c) $t=55$, and (d) $t=127$.}
\label{fig:flux_diss_schem}
\end{figure}

In Fig.~\ref{fig:flux_diss_schem}, we illustrate various energy fluxes for the wavenumber sphere of radius $k=8$ at $t=0.05, 14, 55$, and $127$.  The large-scale magnetic field receives energy mainly via $\Pi^{u>}_{b<}$, and via $\Pi^{b>}_{b<}$ at initial times.  Note that the accumulated energy dissipates minimally due to the $k^2$ factor of $ \eta k^2 E_b(k)$.  Thus, the magnetic energy locked at large-scales remain there for a long time. Noticeably, $\Pi^{b<}_{b>}$ is  positive  after initial transients when it is negative.

In the next subsection we will discuss the shell-to-shell energy transfers during the magnetic energy growth.

\subsection{Shell-to-shell energy transfers}

Energy fluxes provide information about the cumulative energy transfers in wavenumber space. For a refined picture of energy transfers, we compute shell-to-shell energy transfers among velocity and magnetic fields for the wavenumber shells described in Sec.~\ref{sec:simulation}.  The first six wavenumber shells are $(0,2)$, $(2,4)$, $(4,8)$, $(8,9.6)$, $(9.6,11.6)$, and $(11.6,13.9)$.  Since the forcing wavenumber band is $k=$(10--12), it lies in the fifth and sixth shells.  

\begin{figure}[htbp]
\centering
\includegraphics[width=8.5cm,angle=0]{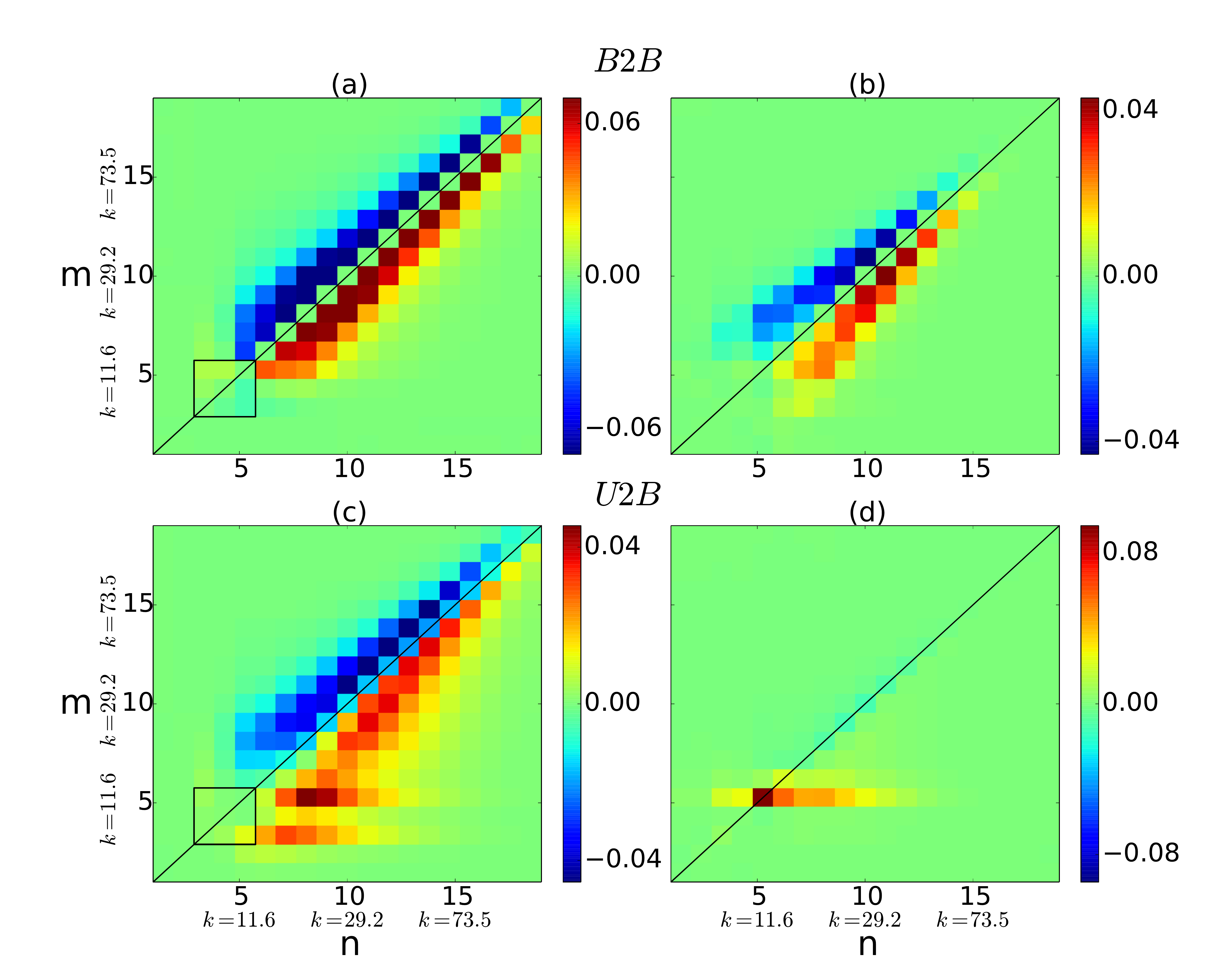} 
\caption{The magnetic to magnetic energy transfers, $B2B$ (top row), and the velocity to magnetic energy transfers, $U2B$ (bottom row) at $t=0.05$ (a, c) and $t=127$ (b, d). The energy transfers corresponding to the boxes of sub-figures (a) and (c) are shown in Fig.~\ref{fig:lsd_shell_zoom}. The vertical axes depict the giver shells, while the horizontal axes represent the receiver shells.}
\label{fig:lsd_shell_uu_bb_ub}
\end{figure}

In Fig.~\ref{fig:lsd_shell_uu_bb_ub}, we illustrate the shell-to-shell energy transfer rates, $B2B$ (magnetic to magnetic) and $U2B$ (velocity to magnetic) at $t=0.05$ (initial phase) and at $t=127$ (final phase).  The indices of the vertical and horizontal axes represent the giver and receiver shells, respectively.  The figure indicates that the  shell $m$ gives energy most dominantly to $(m+1)$, but receives energy from $(m-1)$.  Thus both $B2B$ and $U2B$ energy transfers are local and forward, except those involving the forcing wavenumbers (to be discussed below).  This result is similar to what has been reported by Kumar {\em et al}.~\cite{Kumar:JT2015}  

\begin{figure}[htbp]
\centering
\includegraphics[width=8.5cm,angle=0]{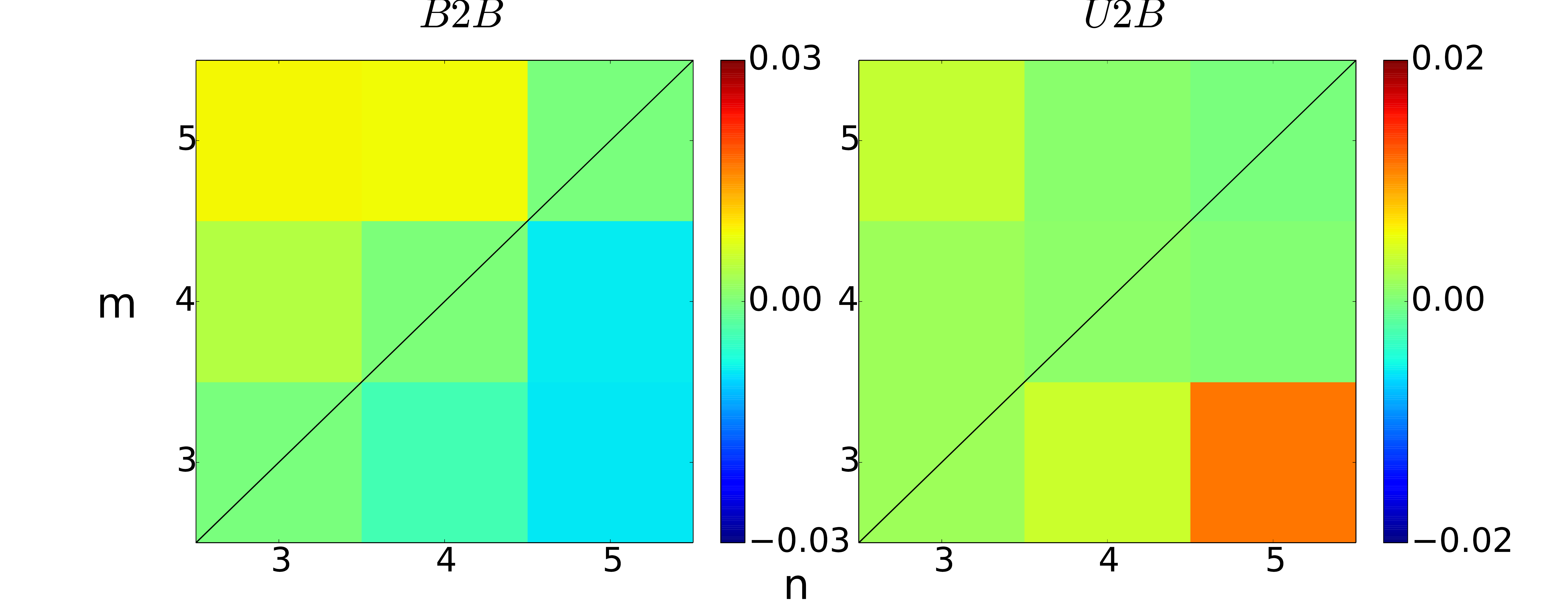} 
\caption{A zoomed views of the shell-to-shell energy transfer rates $B2B$ and $U2B$ at $t=0.05$, corresponding to the boxes of Fig.~\ref{fig:lsd_shell_uu_bb_ub}(a,c).}
\label{fig:lsd_shell_zoom}
\end{figure}

\begin{figure}[htbp]
\centering
\includegraphics[width=8.0cm,angle=0]{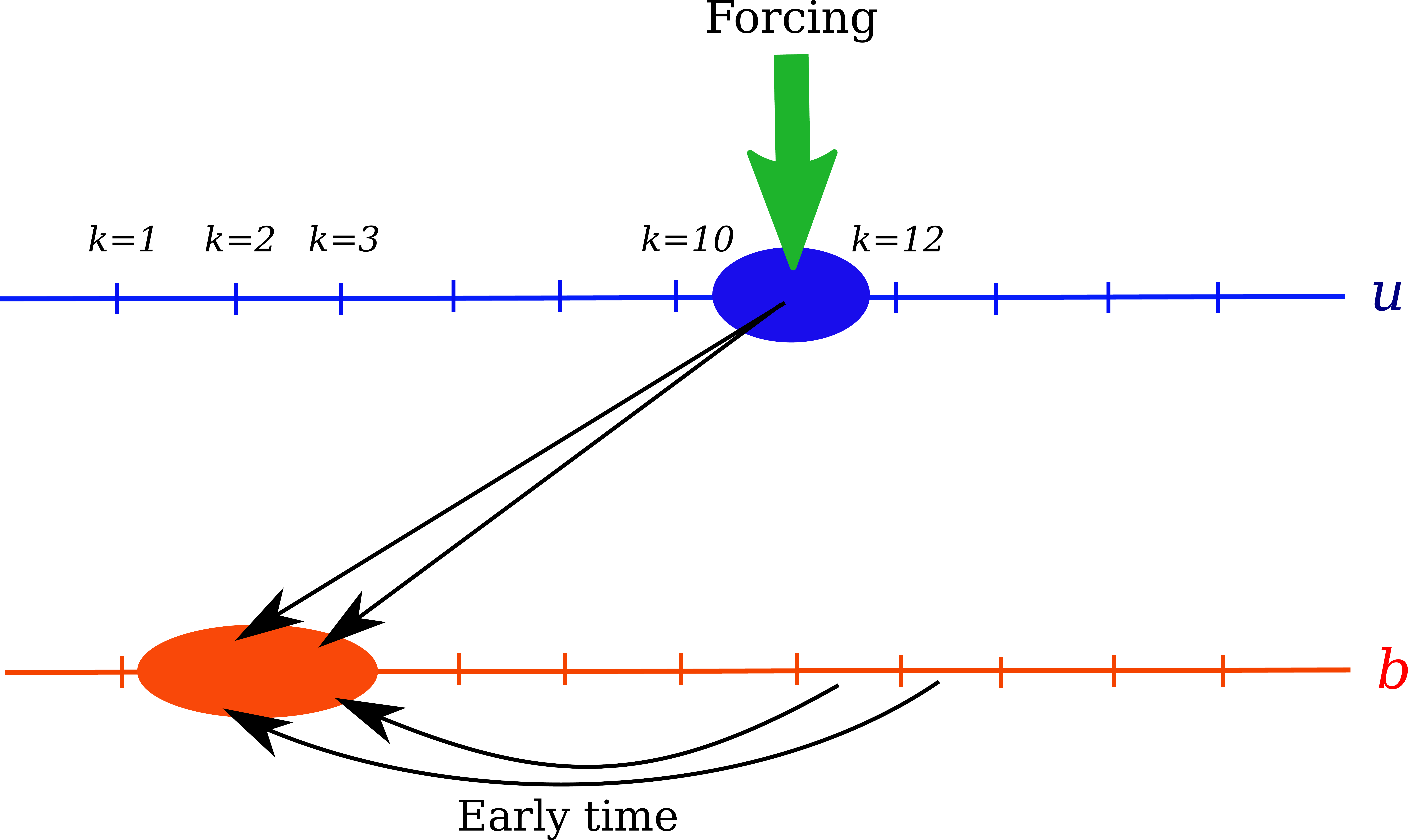} 
\caption{A schematic diagram depicting the dominant energy transfers between velocity and magnetic fields in the wavenumber space at earlier stage, e.g., $t=0.05$.  The velocity and magnetic fields at the forcing wavenumber band transfers energy to the large-scale magnetic field. Note that the inverse  energy transfer of the magnetic energy  takes place only at early stages.}
\label{fig:schem_en}
\end{figure}

However, in the initial stages ($t=0.05$), both $u$ and $b$ fields in the forcing band transfer  energy to the magnetic field at lower wavenumbers.  This feature is clearly visible in the zoomed view of the energy transfers for the shells 3 to 5 (see Fig.~\ref{fig:lsd_shell_zoom}).  The most dominant transfers are from the fifth shell (containing the forcing band) to $n=3$ and $4$.   These inverse energy transfers are responsible for the growth of large-scale magnetic field.  We summarise the above findings in Fig.~\ref{fig:schem_en} that shows energy transfers from $u>$ to $b<$, and $b>$ to $b<$ (in the early phases).  Note however that the inverse transfer of $B2B$ transfers is applicable only in the initial stages.

Another noticeable nonlocal shell-to-shell energy transfer is $U2B$ from the forcing wavenumber band.  The fifth shell of $u$ field transfers energy to the $b$ field of shells 1 to 10 [see Fig.~\ref{fig:lsd_shell_uu_bb_ub}(d)].  These energy transfers are responsible for the strengthening of the magnetic field at large-scales as well as at the intermediate scales.

\section{Conclusions}
\label{sec:conclude}
It is normally assumed  that the kinetic  and/or magnetic helicities are very important for the growth of large-scale magnetic field (or large-scale dynamo) when the forcing is employed at intermediate scales. In this paper we show that large-scale dynamo can occur in nonhelical MHD.  We perform a direct numerical simulation on $512^3$ grid with nonhelical forcing at intermediate wavenumbers $k=[10,12]$. Both kinetic and magnetic helicities are negligible, and the initial magnetic field is present at $k\ge 10$. We observe that the magnetic field at large scale grows due to energy transfers from the forcing wavenumber  band. 

To investigate the growth of large-scale magnetic field, we study the energy flux and shell-to-shell energy transfer.  Our detailed analysis show that the  velocity and magnetic fields at forcing wavenumbers supply energy to the magnetic field at large scales. The aforementioned $B2B$ energy transfer occurs in the initial stage, but $U2B$ energy transfer persists for all the time.  The magnetic energy thus accumulated at large scale is weakly dissipated  due to the $k^2$ factor in the magnetic dissipation $\eta k^2 E_b(k)$.  Also, large-scale magnetic field receives energy from the large-scale velocity field that aids the growth of the large-scale magnetic field. 

Thus, we demonstrate that the large-scale magnetic field can be amplified without kinetic and/or magnetic helicities.  The growth of the magnetic field however will be enhanced in the presence of helicity.  Verma~\cite{Verma:PR2004} observed that the magnetic energy flux due to helicity is
\begin{equation}
\Pi^{b<}_{(b>,u>)\mathrm{helical}} = -a r_M^2 + b r_M r_K
\end{equation}
where $r_K = H_K(k)/(k E_u(k))$ and $r_M = k H_M(k)/E_b(k)$ with $H_K, H_M$ as the kinetic helicity and the magnetic helicity, respectively.  Note that the magnetic energy flux due to helicities is negative (inverse cascade) when  $r_M r_K < 0$.  The aforementioned helical energy flux is in addition to the forward $B2B$ flux.~\cite{Verma:PR2004,Debliquy:PP2005}  Similar observations regarding helicity has been made by Pouquet {\em et al.}~\cite{Pouquet:JFM1976} and Brandenburg.~\cite{Brandenburg:APJ2001}  Thus, the inverse magnetic energy flux induced by helicity aids to the amplification of the large-scale magnetic field.  It will be interesting to extend the analysis of the present paper to helical regime and investigate the large-scale magnetic field. 

In conclusion,  we show that kinetic and magnetic helicities are not absolute requirements for the growth of magnetic field.


\begin{acknowledgments}
We are grateful to the anonymous referee for comments that helped us improve the manuscript. We thank Amitava Bhattacharjee and Rodion Stepanov for the suggestions and comments, and Abhishek Kumar for his help with performing some simulations. The computer simulations were performed on {\em Shaheen II} of the Supercomputing Laboratory at King Abdullah University of Science and Technology (KAUST) under the project K1052, and on {\em Chaos} supercomputer of Simulation and Modeling Laboratory (SML), IIT Kanpur.  This work was supported by the Indo-French research project SERB/F/3279/2013-14 from Science and Engineering Research Board, India and by the Indo-Russian project (DST-RSF) INT/RUS/RSF/P-03 and  RSF-16-41-02012.
\end{acknowledgments}


%


\bibliography{turbulence}

\end{document}